# Edge Intelligence for Wildlife Conservation: Real-Time Hornbill Call Classification Using TinyML

Kong Ka Hing[1], Mehran Behjati[1,2], Vala Saleh[3], Yap Kian Meng[1,2], Anwar P.P. Abdul Majeed[1], Yufan Zheng[4]

[1] Department of Computing and Information Systems, School of Engineering and Technology, Sunway University, 47500, Malaysia
[2] Research Centre for Human-Machine Collaboration (HUMAC), Department of Computing and Information Systems, School of Engineering and Technology, Sunway University, 47500, Malaysia
[3] Department of Electrical Engineering, Shiraz University of Technology, Shiraz, Iran
[4] School of Intelligent Manufacturing Ecosystem, XJTLU Entrepreneur College, Xi'an Jiaotong– Liverpool University (XJTLU), Taicang, Suzhou, 215400, China
mehranb@sunway.edu.my

**Abstract.** Hornbills, an iconic species of Malaysia's biodiversity, face threats from habitat loss, poaching, and environmental changes, necessitating accurate and real-time population monitoring that is traditionally challenging and resource intensive. The emergence of Tiny Machine Learning (TinyML) offers a chance to transform wildlife monitoring by enabling efficient, real-time data analysis directly on edge devices. Addressing the challenge of wildlife conservation, this research paper explores the pivotal role of machine learning, specifically TinyML, in the classification and monitoring of hornbill calls in Malaysia. Leveraging audio data from the Xeno-canto database, the study aims to develop a speech recognition system capable of identifying and classifying hornbill vocalizations. The proposed methodology involves preprocessing the audio data, extracting features using Mel-Frequency Energy (MFE), and deploying the model on an Arduino Nano 33 BLE, which is adept at edge computing. The research encompasses foundational work, including a comprehensive introduction, literature review, and methodology. The model is trained using Edge Impulse and validated through real-world tests, achieving high accuracy in hornbill species identification. The project underscores the potential of TinyML for environmental monitoring and its broader application in ecological conservation efforts, contributing to both the field of TinyML and wildlife conservation.

## 1 Introduction

Hornbills are an iconic and crucial part of Malaysia's biodiversity, playing a vital role in maintaining the ecological balance. However, these magnificent birds face significant threats due to habitat loss, poaching, and environmental changes. Effective con-servation efforts require accurate and real-time monitoring of hornbill populations, which has traditionally been a challenging and resource-intensive task. With the advent of Tiny Machine Learning (TinyML), there is an opportunity to revolutionize wildlife monitoring through efficient and real-time data processing directly on edge devices.

TinyML [1] is a rapidly evolving field that focuses on deploying machine learning models on resource-constrained devices such as microcontrollers. This approach offers numerous advantages, including low power consumption, real-time processing capabilities, and the ability to operate in remote and harsh environments. In this project, we leverage the capabilities of TinyML to develop a system for classifying and monitoring hornbill calls, facilitating conservation efforts in their natural habitats.

The primary objective of this project is to create a robust and efficient system that can accurately classify hornbill calls using audio data. The system is designed to operate on the Arduino Nano 33 BLE Sense, a microcontroller equipped with various sensors and low-power processing capabilities. By utilizing audio data from the Xeno-canto database [2], we aim to train a model that can identify different hornbill species based on their vocalizations. This real-time classification can provide valuable insights into hornbill populations and their behaviors, aiding conservationists in making informed decisions.

The Xeno-canto database is a comprehensive online repository of bird sounds, containing recordings from around the world. For this project, we focus on audio data specific to Malaysian hornbill species. The preprocessing stage involves cleaning the audio data, removing noise, and segmenting the recordings into manageable clips. Feature extraction is performed using Mel-Frequency Energy (MFE), a technique that captures essential characteristics of audio signals and is well-suited for machine learning applications.

The model training process is conducted using the Edge Impulse platform [3], which provides tools for building and optimizing machine learning models for edge devices. The trained model is then deployed on the Arduino Nano 33 BLE Sense, enabling real-time classification of hornbill calls. The system's performance is evaluated based on accuracy, inferencing time, and resource usage, ensuring that it meets the requirements for field deployment.

This paper is structured as follows: Section II reviews related work in the field of wildlife monitoring and the application of TinyML. Section III describes the methodology, including system design, data collection, preprocessing, feature extraction, and model training. Section IV presents the results and discussion, highlighting the system's performance and potential impact on conservation efforts. Finally, Section V concludes the paper and suggests future directions for research.

The implementation of TinyML in hornbill conservation marks a significant advancement in the application of advanced technologies for environmental monitoring. By enabling the real-time classification and monitoring of hornbill calls, this project aims to provide conservationists with a powerful tool to protect these vital birds and preserve Malaysia's rich biodiversity.

## 2 Literature Review

In recent years, the application of machine learning and edge computing technologies in wildlife conservation has gained significant attention. Various studies, such as [4], have demonstrated the effectiveness of these technologies in monitoring and preserving biodiversity. This section reviews the existing literature on audio-based species identification and monitoring systems as well as TinyML applications in wildlife conservation.

One notable contribution in this field is by R.Vishnuvardhan et al. [5], who developed an automatic system for detecting flying bird species using computer vision techniques. They utilized convolutional neural networks (CNNs) to process images captured by a 360-degree camera, achieving notable accuracy in bird species identification. This approach, while effective, relies heavily on visual data and may not be suitable for species that are more easily detected through their vocalizations.

In [6], Mario Lasseck explored the use of deep learning techniques for acoustic bird detection. They designed models to identify 1,500 bird species using Deep Convolutional Neural Networks (DCNNs) pre-trained on ImageNet [7] and subsequently fine-tuned with audio data. Their approach demonstrated the potential of using audio signals for species identification, achieving high accuracy rates in controlled environments.

Stowell et al. [8] utilized machine learning techniques to analyze audio recordings of birds. They emphasized the importance of feature extraction methods, such as Mel-Frequency Cepstral Coefficients (MFCCs), for improving classification accuracy. Their findings suggest that effective feature extraction is crucial for developing robust audio-based monitoring systems.

The authors in [9] present a collaborative data challenge that highlights the effectiveness of modern machine learning, particularly deep learning, in the field of acoustic bird detection. The challenge demonstrated that general-purpose systems could achieve high retrieval rates in remote monitoring data without the need for manual recalibration or pre-training for specific bird species or acoustic environments. The introduction of new acoustic monitoring datasets highlighted the diverse machine learning techniques employed by participating teams, accompanied by a thorough performance evaluation. The authors also discuss the potential integration of these detection methods into remote monitoring projects, highlighting the practical applications of their findings.

Kahl et al. [10] demonstrated the use of Convolutional Recurrent Neural-Networks (CRNNs) for bird audio detection. By combining convolutional layers for feature extraction and recurrent layers for temporal modeling, their approach showed improved performance over traditional Convolutional Neural-Networks (CNNs), particularly in handling sequential audio data.

In [11], Knight et al. highlight the challenges and opportunities in deploying edge Artificial Intelligence (AI) for environmental monitoring. They discussed the practical considerations for implementing machine learning models on low-power devices, including the need for efficient model compression techniques and the trade-offs between accuracy and computational complexity.

In another relevant study, Grillo et al. [12] applied machine learning to classify frog calls using edge devices. Their work demonstrated the feasibility of deploying audio-based species identification systems in the field, highlighting the potential for similar applications in bird conservation.

Bergler et al. [13] reviewed the advancements in TinyML and its applications across various domains. They emphasized the potential of TinyML in enabling real-time, on-device inference, which is particularly valuable for continuous wildlife monitoring in remote areas.

In [14], Chowdhury et al. investigated the potential of TinyML for deploying machine learning algorithms on resource-constrained devices in the context of wildlife conservation. The study examined various case studies, offering detailed insights into the practical challenges and advantages of utilizing TinyML for species identification and environmental monitoring.

Recent literature underscores that the integration of TinyML in wildlife conservation represents a cutting-edge advancement in the field. TinyML facilitates the deployment of machine learning models on microcontrollers and other resource-constrained devices, making it particularly suited for remote and real-time monitoring applications [15]. Research, such as that conducted by Edge Impulse, has demonstrated the feasibility of deploying TinyML models across various contexts, including environmental monitoring and species identification. These studies emphasize critical advantages, including low power consumption, real-time data processing, and the ability to function effectively in remote locations without continuous internet connectivity.

In summary, the existing literature highlights the substantial potential of machine learning and edge computing technologies in wildlife conservation. Table I provides a summary of the reviewed studies. However, there remains a notable gap in research specifically focused on audio-based species identification using TinyML. This project aims to address this gap by developing a comprehensive solution for hornbill conservation in their natural habitats, utilizing TinyML to enhance the effectiveness and efficiency of species identification and monitoring.

**Table 1.** Summary of Reviewed Studies.

| Study | Methodology | Application |
|---|---|---|
| Vishnuvardhan et al. [5] | CNN, 360-degree camera | Bird species detection |
| Lasseck et al. [6] | DCNNs, audio data | Acoustic bird detection |
| Stowell et al. [8] | MFCCs, machine learning | Bird audio monitoring |
| Stowell et al. [9] | Deep learning, Post-processing | Bird detection |
| Kahl et al. [10] | CRNNs | Bird audio detection |
| Knight et al. [11] | Edge AI, model compression | Environmental monitoring |
| Grillo et al. [12] | Machine learning, edge devices | Frog call classification |
| Bergler et al. [13] | TinyML, on-device inference | Various domains |
| Chowdhury et al. [14] | TinyML, case studies | Wildlife conservation |

## 3 Methodology

This section outlines the comprehensive methodology adopted for the implementation of a TinyML-based system aimed at the conservation of hornbills in their natural habitats. The primary objective of this project is to develop a robust and efficient machine learning model that can accurately classify hornbill calls. This model has been deployed on an Arduino Nano 33 BLE Sense for real-time monitoring and conservation efforts. The methodology is systematically structured to ensure that each phase of the project is addressed, from the initial system design to the final deployment in the field. Fig. 1 illustrates the steps of the proposed methodology detailed in this research.

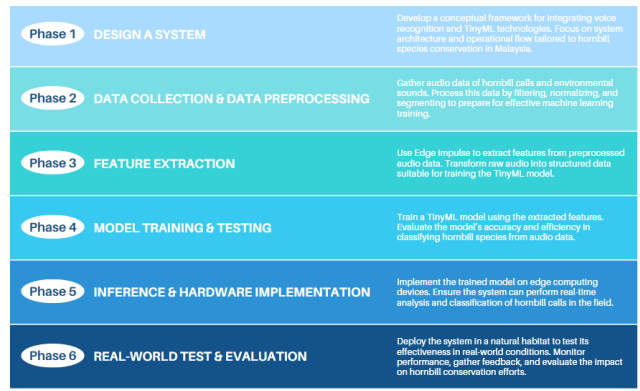

**Fig. 1.** Overall diagram of the proposed methodology.

### 3.1 System Architecture

The proposed system is designed to integrate voice recognition with TinyML to accurately classify hornbill calls. The system's architecture revolves around the utilization of lightweight machine learning models compatible with low-power hardware, facilitating on-site data processing and analysis. The objective is to provide real-time analysis and monitoring capabilities, enabling timely and effective conservation interventions.

The central hardware component selected for this system is the Arduino Nano 33 BLE Sense (Fig. 2), chosen for its robust features, including Bluetooth connectivity, low energy consumption, and support for on-device machine learning tasks.

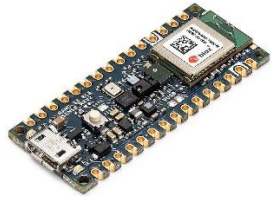

**Fig. 2.** Arduino Nano 33 BLE Sense.

The software stack for the system is built around Edge Impulse, a versatile platform designed for developing, deploying, and managing machine learning models on edge devices.

### 3.2 Data Collection & Data Preprocessing

Data collection is a critical phase for developing a robust machine learning model capable of accurately classifying hornbill calls. Audio recordings of five hornbill species (Black Hornbill, Bushy-Crested Hornbill, Oriental Pied Hornbill, Rhinoceros Hornbill, and White-Crowned Hornbill) were sourced from Xeno-Canto.
Preprocessing involves several critical steps to enhance the quality of the audio data: Noise reduction was performed using Audacity [16] to minimize background noise, ensuring the primary focus remained on the hornbill vocalizations (Fig. 3).

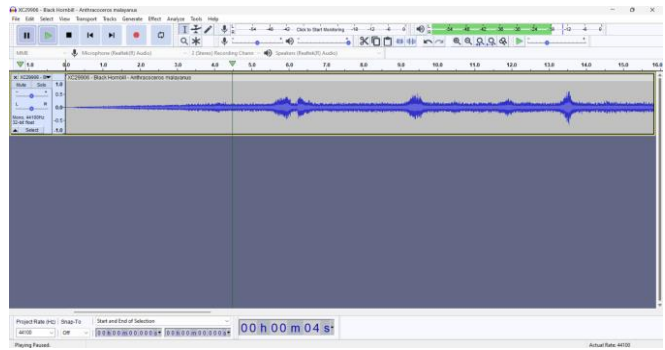

**Fig. 3.** Diagram of the software of Audacity.

Normalization adjusted the amplitude of the audio recordings to a consistent range. Segmentation divided each recording into 10-second segments to increase the diversity of the training data (Fig. 4).

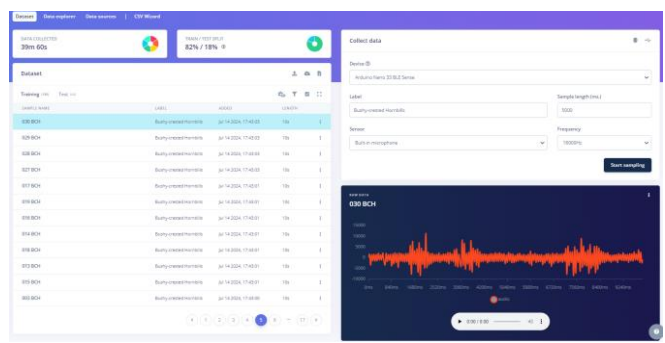

**Fig. 4.** Balanced datasets collected in Edge Impulse.

### 3.3 Feature Extraction

Feature extraction transforms raw audio data into a structured format suitable for machine learning. This process, conducted using Edge Impulse, involves several key stages. First, **windowing** segments the audio stream into smaller, overlapping windows to capture its temporal characteristics of the audio signal. Selecting blocks involves choosing the Audio Processing Block (MFE) and the Classification Learning Block. Configuring MFE blocks includes setting parameters such as frame length, frame stride, Mel filters, FFT length, and low frequency cutoff.

Normalization standardizes the input to ensure consistency across the dataset, with the noise floor set to -52 dB to filter out low-level background noise. Digital Signal Processing (DSP) results, including Mel Energies and FFT Bin Weighting, provide a detailed understanding of the audio signal's structure and features, aiding in the classification of hornbill calls (Fig. 5).

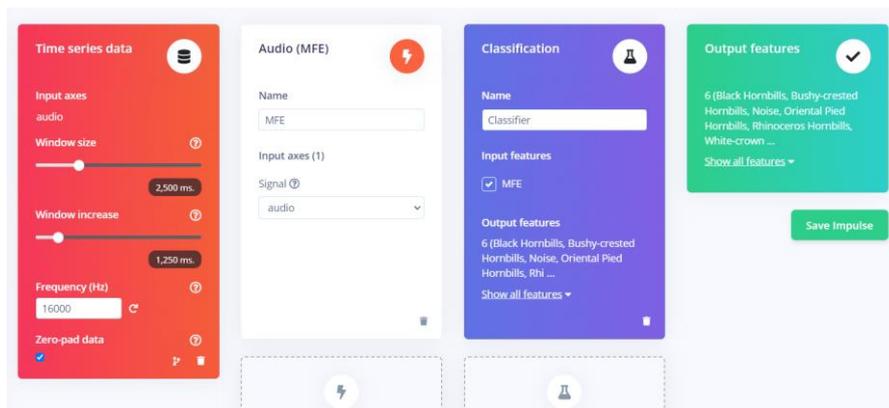

**Fig. 5.** Configuration of feature extraction in Edge Impulse.

### 3.4 Model Training & Testing

The implemented Artificial Neural Network (ANN) architecture in Edge Impulse was optimized to classify hornbill audio signals. The model architecture begins with an input layer of 7,470 features extracted from the MFE process, reshaped into 30 columns. These features are processed through two one-dimensional convolutional layers with a kernel size of 3 and ReLU activation functions to capture complex patterns in the audio data. Each convolutional layer includes a dropout rate of 0.25 to mitigate overfitting.
A flatten layer follows, converting the multi-dimensional outputs into a single vector. The output layer comprises six classes representing five hornbill species and noise, utilizing a softmax activation function for classification. The model was trained with a learning rate of 0.005 over 100 epochs, achieving an accuracy of 97.5% with k-fold cross-validation ensuring robustness.

Rigorous testing was conducted using the Live Classification feature in Edge Impulse on both smartphones and the Arduino Nano 33 BLE Sense, demonstrating efficient real-time classification and validating the model's deployment capability on low-power edge devices. This approach confirms the model's suitability for practical wildlife conservation efforts. Fig. 6 depicts the configurations of ANN in Edge Impulse.

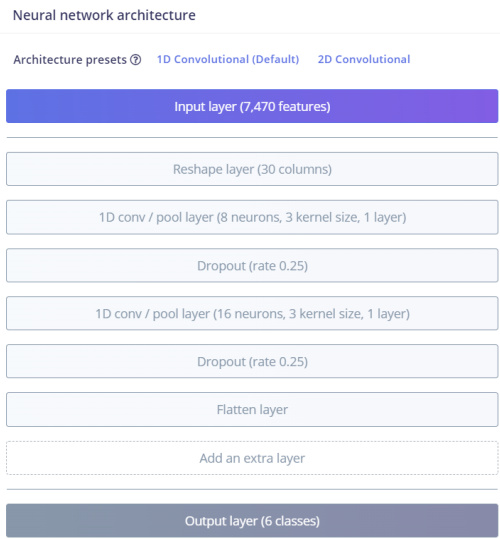

**Fig. 6.** Configuration of neural network architecture in Edge Impulse.

### 3.5     Inference & Hardware Implementation

The trained model is optimized through techniques such as compression and quantization for deployment on the Arduino Nano 33 BLE Sense. Edge Impulse generates a custom Arduino library that includes the optimized model and all necessary dependencies. This library is integrated into the Arduino Integrated Development Environment (IDE), which initializes the microcontroller and connects to the microphone for real-time data acquisition and inference.

### 3.6     Real-world Test & Evaluation

The Edge Impulse library, containing the trained ML model, is installed in the Arduino IDE with configuration parameters aligned to those used during the training phase. Fig. 7 illustrates the installation process of the library. The compiled firmware is then uploaded to the Arduino Nano 33 BLE Sense, enabling the execution of the machine learning model. Once the code is successfully uploaded, the system can be deployed in the natural habitats of hornbills, where it captures and classifies hornbill calls in real-time. Performance metrics, including detection accuracy, false positive rates, and response times, are recorded to assess the system's effectiveness.

**Fig. 7.** Compiling the code to Arduino Nano 33 BLE Sense.

## 4 Results

### 4.1 Feature Explorer After Generating Features

The feature explorer provides a visual representation of the extracted features and their distribution across different classes. Using MFE extraction method, the feature explorer offers insights into the separability and clustering of the hornbill call features. The feature explorer diagram shown in Fig. 8 illustrates the distribution of audio features across six classes: Black Hornbills, Bushy-crested Hornbills, Noise, Oriental Pied Hornbills, Rhinoceros Hornbills, and White-crowned Hornbills. Each point represents a 2.5-second audio window processed by the MFE algorithm, plotted in a two-dimensional feature space. The points are color-coded according to their respective classes, facilitating an understanding of feature separability. Distinct clusters for each class indicate effective feature extraction, essential for accurate classification by the machine learning model.

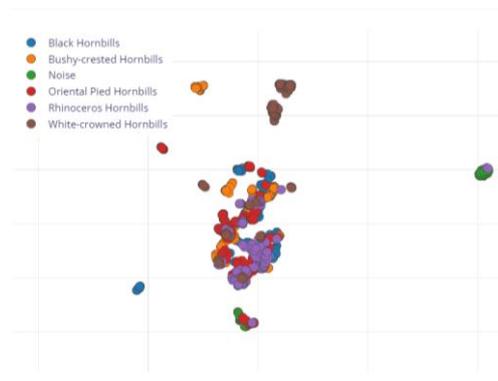

**Fig. 8.** Diagram of Feature Explorer.

## 4.2 Model Training Performance

The performance of the model during training is crucial for understanding its ability to learn and generalize from the data. The model achieved an accuracy of 97.5% on the validation set, with a corresponding loss of 0.19 (Fig. 9). These metrics indicate the model's capability to classify hornbill calls accurately.

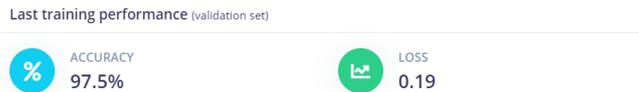

**Fig. 9.** Training performance of model training.

The confusion matrix provides a detailed breakdown of the model's performance across different classes, showing high true positive rates and minimal misclassifications. The model exhibited high accuracy for all hornbill species and noise, reflecting its robustness in distinguishing between classes (Fig. 10).

| | BLACK HORNB | BUSHY-CRESTE | NOISE | ORIENTAL PIED | RHINOCEROS F | WHITE-CROWN |
|---|---|---|---|---|---|---|
| BLACK HORNBIL | 92.9% | 0% | 2.4% | 0% | 2.4% | 2.4% |
| BUSHY-CRESTED | 2.2% | 97.8% | 0% | 0% | 0% | 0% |
| NOISE | 0% | 0% | 100% | 0% | 0% | 0% |
| ORIENTAL PIED | 0% | 2.9% | 0% | 97.1% | 0% | 0% |
| RHINOCEROS HC | 2.1% | 0% | 0% | 0% | 97.9% | 0% |
| WHITE-CROWNE | 2% | 0% | 0% | 0% | 0% | 98% |
| F1 SCORE | 0.93 | 0.98 | 0.99 | 0.99 | 0.98 | 0.98 |

**Fig. 10.** Confusion matrix of model training.

## 4.3 Model Deployment to Device

The deployment of the trained model to the Arduino Nano 33 BLE Sense was optimized using the EON™ Compiler, reducing the model's memory footprint and computational requirements. The total latency for running the model, including MFE feature extraction and classification, is reduced to 484 milliseconds (Fig. 11).
The on-device performance metrics confirm that the model operates efficiently within the constraints of the Arduino Nano 33 BLE Sense. The optimized model requires less memory and computational power, ensuring efficient and timely predictions.

11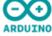

**Fig. 11.** Default deployment and model optimizations.

### 4.4 Real-Time Prediction Rates Using Arduino Nano 33 BLE Sense

The final evaluation phase involved testing the real-time prediction rates of the trained model deployed on the Arduino Nano 33 BLE Sense (Fig. 12). The device processed audio data and made real-time predictions with high accuracy. The real-time performance metrics, including DSP time and classification time, demonstrate the system's ability to process and classify audio data efficiently. The high confidence scores in real-time predictions validate the model's robustness and reliability, ensuring accurate classification of hornbill calls in practical scenarios.

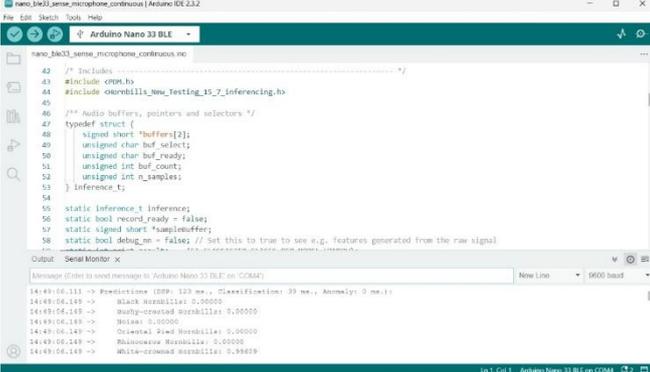

**Fig. 12.** Diagram of real-time prediction rates of the trained model deployed on the Arduino device.

12## 5      Conclusion

The study demonstrates the practical viability and effectiveness of a TinyML-based system for hornbill conservation. The feature extraction process, utilizing the Mel Frequency Cepstral Coefficients method, successfully captured critical characteristics of hornbill calls, enabling precise species classification. The trained model achieved high accuracy and robust performance metrics, affirming its capability to differentiate between various hornbill species and background noise.

Deployment of the model on the Arduino Nano 33 BLE Sense, optimized with the EON™ Compiler, confirmed the feasibility of using resource-constrained devices for real-time monitoring. The system's real-time prediction capabilities ensure immediate detection and classification of hornbill calls, which is essential for timely conservation interventions.

In conclusion, this TinyML-based system represents a significant advancement in wildlife monitoring and conservation. By integrating advanced machine learning techniques with edge computing, the system offers a valuable tool for real-time, on-site analysis of hornbill vocalizations. This approach not only enhances the monitoring and protection of these vital bird species but also underscores the broader potential of deploying machine learning models on low-power devices for environmental conservation efforts.

## References

1. J. Lin, L. Zhu, W. -M. Chen, W. -C. Wang and S. Han, "Tiny Machine Learning: Progress and Futures [Feature]," in IEEE Circuits and Systems Magazine, vol. 23, no. 3, pp. 8-34, thirdquarter 2023
2. Xeno-Canto Homepage, https://xeno-canto.org, last accessed 2024/07/31.
3. Edge Impulse Homepage, https://edgeimpulse.com/, last accessed 2024/07/31.
4. Zeyu Xu, Tiejun Wang, Andrew K. Skidmore, Richard Lamprey, A review of deep learning techniques for detecting animals in aerial and satellite images, International Journal of Applied Earth Observation and Geoinformation, Volume 128, 2024, 103732, ISSN 1569-8432.
5. R. Vishnuvardhan et al., "Automatic detection of flying bird species using computer vision techniques , 2019 J. Phys.: Conf. Ser. 1362 012112.
6. M. Lasseck, "Acoustic Bird Detection with Deep Convolutional Neural Networks,Published in Workshop on Detection and Classification of Acoustic Scenes and Events 2018, Computer Science, Environmental Science.
7. Deng J, Dong W, Socher R, Li LJ, Li K, Fei -Fei L (2009) Imagenet: A largescale hierarchical image database. In: IEEE Conference on Computer Vision and Pattern Recognition, 2009. pp. 248–255.
8. Stowell, D., Giannoulis, D., Benetos, E., Lagrange, M. and Plumbley, M. D. (2015) Detection and classification of acoustic scenes and events. IEEE Transactions on Multimedia, 17, 1733–1746.
9. D. Stowell et al., "Automatic acoustic detection of birds through deep learning: The first Bird Audio Detection challenge, Methods in Ecology and Evolution, Wiley ,2018, vol 10, 368 - 380.